\documentclass{aa}
\usepackage{graphics}

\begin{document}
\thesaurus{10(10.03.1; 11.14.1; 13.09.1; 13.09.6)}

%\noindent
\title{ The Sgr~A$^*$ Stellar Cluster:
\\
 New NIR Imaging and Spectroscopy }

\author{Andreas Eckart
\and Thomas Ott
\and Reinhard Genzel}

\offprints{A. Eckart}

\institute{Max-Planck-Institut f\"ur extraterrestrische Physik, Giessenbachstr.,
D-85740 Garching}

\date{Received / Accepted}

\maketitle

\begin{abstract}
We report preliminary results based on new near-infrared
 observations of the central stellar cluster of 
our Galaxy conducted with the infrared spectrometer ISAAC at the 
ESO VLT  UT1 and the MPE speckle camera SHARP at the ESO  NTT.
We obtained $\lambda$/$\Delta \lambda$$\ge$ 
3000 K-band spectroscopy of the 2.058 $\mu$m He~I, 
2.165$\mu$m Br$\gamma$ emission lines,
and 2.29 $\mu$m CO bandhead absorption line.
These data demonstrate clearly that there is no strong CO bandhead absorption 
originating in the northern part (S1/S2 area) of the central stellar cluster
at the position of Sgr~A$^*$. This makes it likely that these K$\sim$14.5 stars
are
O9 - B0.5 stars with masses of 15 to 20 M$_{\odot}$.
Weaker CO bandhead absorption in the southern part of the cluster (S10/S11 area) could be due to contributions from 
neighbouring stars.
We also report the detection of Br$\gamma$ line emission at the position of the 
central stellar cluster which could be associated with the 'mini-spiral'
rather than with the Sgr~A$^*$ cluster itself.
At the NTT we obtained another epoch epoch of proper motion measurments.
The changes of source positions are consistent with the proper motion
velocities derived from earlier epochs.

\end{abstract}

\section{Introduction}

Recent investigations of the motions of gas and stars have provided 
evidence for the existence of massive black holes in the nuclei of many galaxies
(Richstone et al.  1998, Magorrian et al.  1998, Kormendy and  Richstone 1995).
One of the best cases is the center of our Galaxy itself. 
There both the gas and  stellar dynamics  indicate the
presence of  a large  unresolved central  mass
( Eckart and  Genzel 1996, 1997, Genzel et al.  1997, Ghez et al.  1998, 
Genzel et al. 1999). 
At its measured mass and density it cannot be 
stable and therefore is most likely present
in the form of a massive black hole (Maoz 1998).
The proximity of only 8~kpc to the Galactic Center allows
us to obtain line-of-sight  velocities (through spectroscopy) and/or 
proper motions of individual stars that are within only a few light days of the 
radio/near-infrared  position of Sgr~A$^*$   (Menten et al. 1997).
The positions of the maximum velocity dispersion
and of the maximum stellar  surface density agree with  the position of  the
compact radio source Sgr~A$^*$ to within $\pm$0.1'' (Ghez et al. 1998).
Combined with stellar surface density counts 
these data provide a convincing qualitative evidence for the 
presence of a central point mass ranging between
2.2 and 3$\times$10$^6$ M$_{\odot}$
(Sellgren et al.  1990, Krabbe et al. 1995, Haller et al. 1996, Genzel
et al. 1996, 1997, Eckart and Genzel 1996, 1997, Ghez et al. 1998, Genzel 1999).

Genzel et al. (1997) reported
first R=$\lambda$/$\Delta \lambda$$\sim$35 speckle 
spectroscopy measurements on individual objects in 
the central $\sim$1'' diameter
stellar cluster at the position of Sgr~A$^*$(IR).
In combination with other data this spectroscopic information can be used to
derive a lower limit to the mass associated with the compact radio source.
Here we present very first ISAAC R$\ge$3000  
K-band spectroscopy of the Sgr~A$^*$ stellar cluster in the 
2.058 $\mu$m He~I, 2.165$\mu$m Br$\gamma$
emission lines, and the 2.29 $\mu$m CO bandhead absorption lines.
The combination of these spectroscopic data taken in excellent seeing
(0.3'' to 0.5'') and our new speckle image reconstructions
based on SHARP  NTT data strengthen the case for a compact mass
and add to our understanding of the stellar population near the 
center of the Galaxy.

\section{Observations and Data Reduction}

The spectroscopic observations were carried out in the first half nights of 
30 June and 1 July, 1999, using the infrared (0.9-5$\mu$m) spectrometer
ISAAC (Moorwood et al. 1998) at one Nasmyth focus of the ESO VLT UT1 (ANTU).
The conditions during the first and most of the second 
half night were photometric with a relative humidity below 10\%-20\%,
and low wind from northerly directions.
The seeing indicated by the optical seeing monitor was always between 
0.4'' and 0.6'' resulting an infrared seeing than was always better that 0.5''
and in part $\sim$0.3''.
The very good seeing conditions allowed us to use a 0.6'' slit. 
We observed with two different slit position angles such that
always both the central stellar cluster at the position of Sgr~A$^*$
and one of the neighboring bright sources IRS~16~NW (at a PA of 0$^o$ N-E)
or the bright star between IRS~16~CC and IRS~16~NW 
(at a PA of 108$^o$ N-E) were along the slit.
In all acquisition images 
(2 seconds exposures through a 1\% band filter centered at 2.09$\mu$m)
the small stellar cluster surrounding Sgr~A$^*$  
could easily be identified and was positioned on the slit.
Between the exposures 
the slit position was checked on the direct acquisition images and/or on the 
two dimensional spectroscopic frames.
For each of the three lines (2.058~$\mu$m He~I, 2.165~$\mu$m Br$\gamma$,
 and 2.29~$\mu$m CO(2-0))
we obtained a total of about 8 times 5 minutes integration time per slit setting.
Since the field is very crowded with stars we took in addition to
other calibration data separate sky exposures
on a dark cloud 713 W and 400 N of the central position.
We subtracted darks from all exposures, applied a flatfield, 
conducted a sky-subtraction allowing
for small pixel shifts and a scaling factor to achieve optimum
sky line subtraction.
The wavelength calibration was done using lamp exposures and the 
OH sky lines.
As a spectroscopic reference we used the He~I star IRS~16NE for the CO bandhead
measurements and the late type star 12.05'' north of IRS~7 
for the Br$\gamma$ and He~I measurements. These
stars are bright and featureless in the corresponding wavelength domains.

The diffraction limited imaging data were obtained using the MPE speckle camera SHARP 
at the ESO NTT between 18 to 21 June. On the night from the 19 to 29 June the seeing 
conditions were excellent. For most of the night the optical seeing monitor indicated 
seeing values below 0.5''.
During the first and third night the seeing ranged between 0.5'' and 1''. 
We collected several thousand short exposure frames (0.3 to 0.5 
seconds integration time). The data were then processed in the standard manner 
(dead pixel correction, sky subtraction, flatfielding etc.). 
We then co-add the 256$^2$ pixel frames using the brightest pixel 
in the seeing disks of IRS7 or IRS16 NE as shift-and-add reference.
This is followed by CLEANing  the raw shift-and-add images with the Lucy-Richardson 
(Lucy 1974) algorithm 
%(or a Wiener deconvolution; see Ott et al. 1999 for details), 
using IRS7/IRS16 NE as a point spread function.
We finally re-convolve the resulting maps with a Gaussian restoring beam, 
which had a FWHM near the diffraction limit of the telescope.

\section{Preliminary Spectroscopic Results}

The goal of the spectroscopic observations was 
to obtain high spectral resolution information on the
stars in the central arcsecond. The problem is, that very good 
seeing is required in order
to separate the small flux contribution of the typically K=14.5-15.0 sources from the very bright
neighboring IRS~16 complex which contains stars as bright as K=10.
The excellent seeing conditions we encountered at the VLT allowed us to obtain this information
basically without having to correct for the seeing wing contribution of these bright neighboring
objects. Here we concentrate on two preliminary results: the confirmation of the absence of strong
CO bandhead absorption in the central stellar cluster and the Br$\gamma$ emission we found
towards this region.

\subsection{CO bandhead absorption}
In Fig.1 we compare our new June 1999 speckle image reconstruction 
with a section of the two dimensional ISAAC spectroscopic exposure
on the CO(2-0) bandhead absorption line. This comparison clearly
demonstrates the excellent seeing we had at the VLT and allows us to 
identify the individual sources that contributed to the flux density
in the 0.6'' slit. One can  distinguish between the northern and
southern part of the Sgr~A$^*$ stellar cluster as well as a star 
to the south with obvious bandhead absorption.
In Fig.2 we show spectra of the northern and southern part of the
Sgr~A$^*$ cluster as well as the spectrum of a star just 1.12'' south  
of the central position. The spectra clearly show the complete
absence of strong
CO bandhead absorption for the northern S-sources close to the position of
Sgr~A$^*$ and the detection of a late type star just $\sim$0.6'' south of
the S10 and S11 (Genzel et al. 1997).
Very weak bandhead absorption on the northern Sgr~A$^*$ cluster is in 
agreement with an expected contribution from the underlying stellar cluster.
The bandhead absorption on the southern part of the
central stellar cluster is probably due to a significant flux density 
contribution from the late type star just 0.6'' to the 
south and 0.5'' to the east of S10 and S11. In addition there are a few 
weaker sources with separations from S10 and S11 of less than 0.4''
that could give rise to contaminating flux. 
These measurements are in full agreement with our initial results
that we obtained via R$\sim$~35 speckle spectroscopy 
measurements on the individual objects S1, S2, S8, and S11.
It also indicates that most of the other S-sources that now 
fell in our slit can not be stars with strong CO bandhead absorption.

\begin{figure*}
\resizebox{11.3cm}{!}{\includegraphics{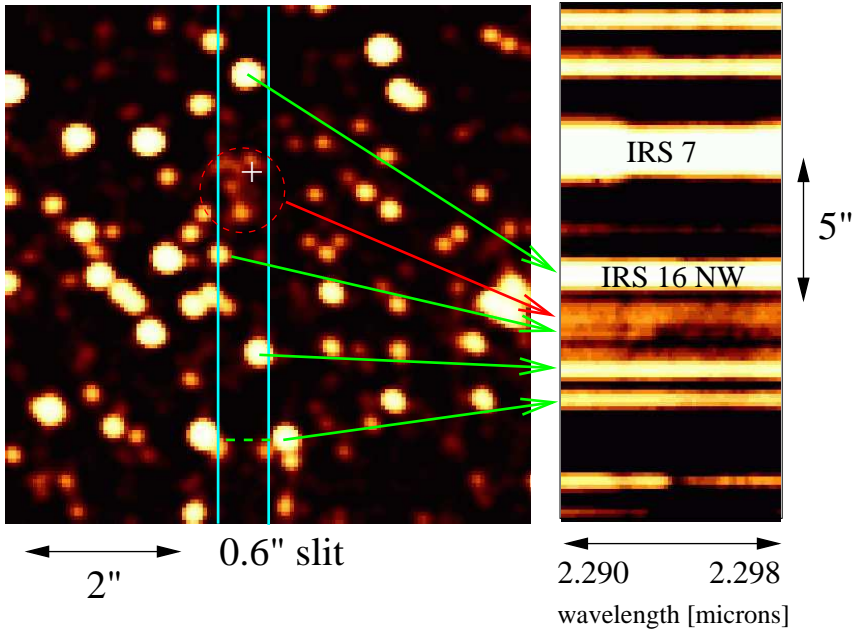}}
\hfill
\caption{
Comparison 
between the results of  new June 1999 SHARP epoch at the NTT (left)
and the central portion of a two dimensional ISAAC spectroscopic exposure
on the CO(2-0) bandhead absorption line (right). 
We indicate the position of the 0.6'' slit and some of the sources that 
contribute to the flux density measured through it. The infrared 
seeing during the 5 minute ISAAC exposure was of the order of 0.3''.
One can clearly distinguish between the northern and southern part 
of the small stellar cluster surrounding the position of Sgr~A$^*$ indicated
by a cross in the SHARP image. 
}
\label{fig1}
\end{figure*}

\begin{figure}
\resizebox{\hsize}{!}{\includegraphics{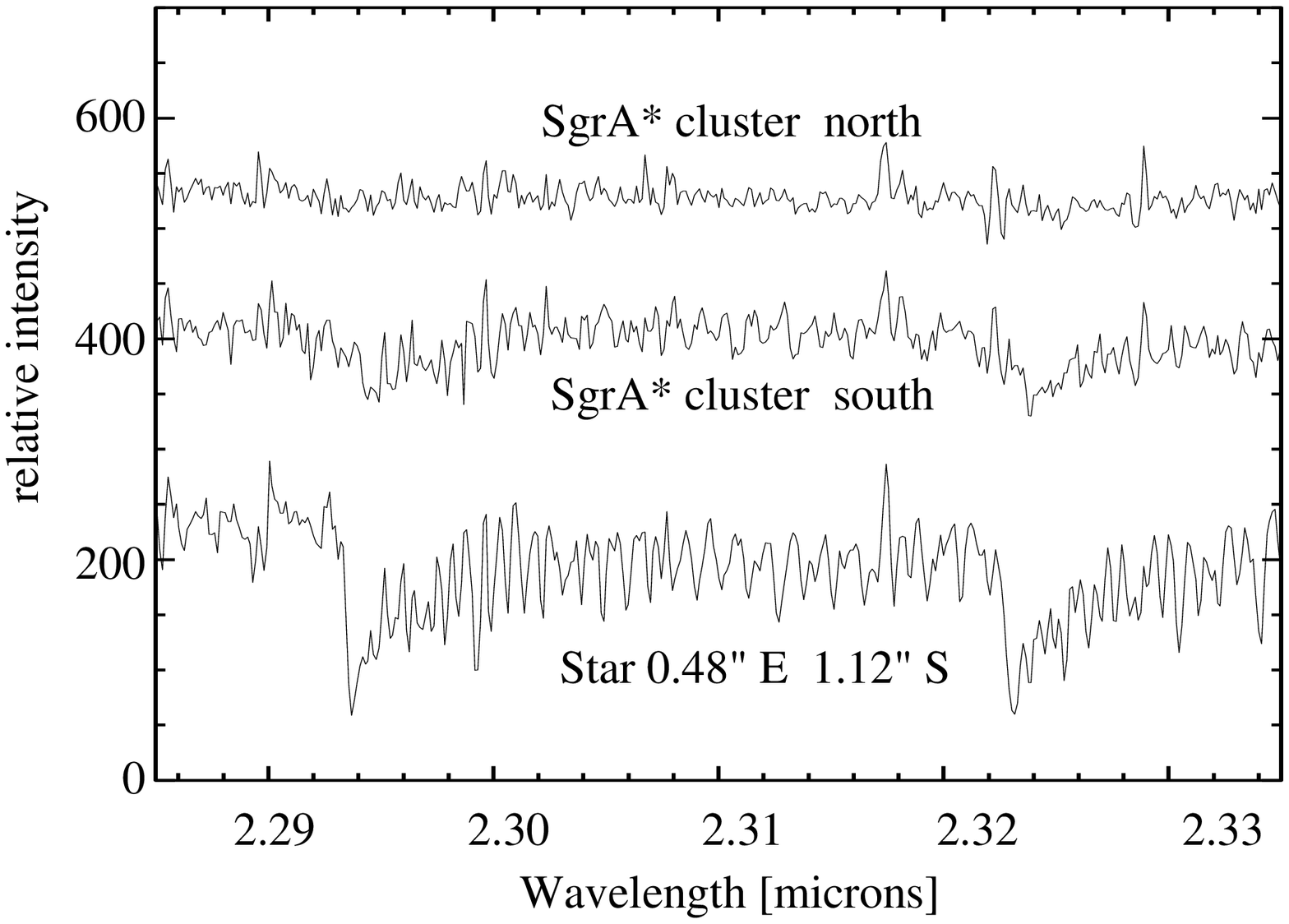}}
\hfill
\caption{
Spectra of the northern and southern part 
of the Sgr~A$^*$ cluster as well as the late type
 star 1.12'' south of the center. 
No CO(2-0) and CO(3-1) bandhead absorption is measured towards the 
northern part containing the fast moving sources S1 and S2.
The weak absorption features towards the south could be due to contaminating
flux density contributions from sources other than S9, S10 and S11 (see text).
The spectrum of the southern part has been shifted down by 150 units;
the spectrum of the late type star was shifted down by 300 units.
}
\label{fig2}
\end{figure}

From this data one can conclude that 
the m$_K$$\sim$14.5 sources in the 
central Sgr~A$^*$  cluster are most likely moderately luminous 
(L$\sim$5,000 to 10,000 L$_{\odot}$) early type stars. If these objects are on the main 
sequence they would have to be O9 - B0.5 stars with masses of 15 to 20 M$_{\odot}$.

Backer (1996) and Reid et al. (1999) have shown that the proper
motion of Sgr~A$^*$ itself is $\le$ 16-20~km/s 
which is close to 2 orders of magnitudes 
smaller than the velocity of the fast moving stars in its vicinity. 
N-body simulations using 20 M$_{\odot}$ as an upper limit of the 
mass distribution 
of these high velocity stars result in a lower
limit of 10$^3$ M$_{\odot}$ for Sgr~A$^*$
(Reid et al. 1999, see also Genzel et al. 1999, 1997).
If this mass is enclosed
within the radio size of Sgr~A$^*$ ($\le$ 1~AU) this 
already implies a central mass
density larger than 10$^{18}$ M$_{\odot}$/pc$^{-3}$.

\subsection{Br$\gamma$ and He~I emission}

Our high spatial and spectral resolution data clearly show the presence of 
Br$\gamma$ and He~I emission which is apparently spatially coincident with the 
location of the Sgr~A$^*$ central stellar cluster.
From our Br$\gamma$ data (see Fig.3) 
we find a line width of $<$120~km/s and  a velocity 
gradient of about 35~km/s between the southern part (S10/S12-region) and the 
northern part (S1/S2-region) of the cluster.
In both slit settings 
this line emission appears to be connected 
to the more extended line 
emission over the remaining central cluster. 
This fact combined with the 
small line width at any position in that region
indicates that the emitting gas is not necessarily associated 
with the Sgr~A$^*$ stellar cluster.
If the emission would be associated with the cluster we would expect a larger 
line width due to the higher gravitational potential indicated by the 
rapid motions of the stars.  However, we cannot exclude at the present 
stage of the data reduction any broad and weak emission components
that would indicate a higher velocity dispersion.

\begin{figure}
\resizebox{\hsize}{!}{\includegraphics{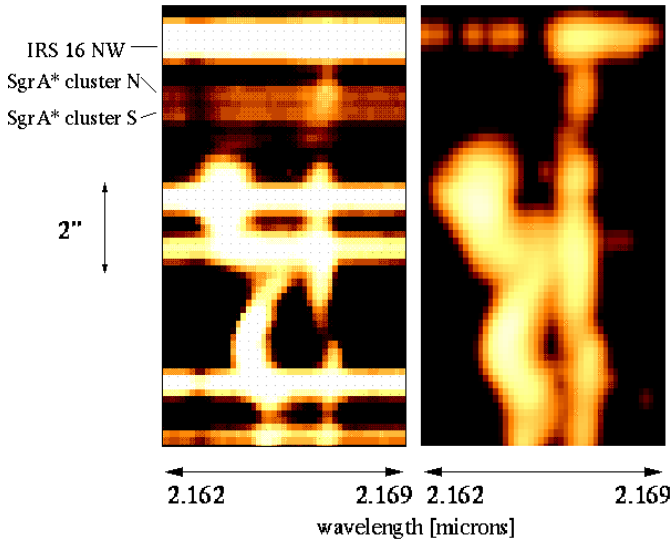}}
\hfill
\caption{
The central portion of a 5 minute ISAAC exposure centered on the Br$\gamma$ 
emission line. On the left we show the line plus continuum map - 
on the right the continuum subtracted wavelength position map.
Some sources are indicated on the left. 
There is a clear detection of Br$\gamma$ emission towards the 
central stellar cluster at the position of Sgr~A$^*$.
At the top of the right panel the 
line emission of IRS~16~NW is shown. The lower wavelength part of the spectrum
includes some residuals from the continuum subtraction.
}
\label{fig3}
\end{figure}

\section{First 1999 Speckle Imaging Results}

Only about 10 days before our spectroscopic measurements we obtained 
new diffraction limited speckle imaging data at the NTT. These data mainly
serve as a new proper motion epoch (Genzel et al.  1997, Eckart and  
Genzel 1996, 1997, Ghez 1998) but also allows us to further
investigate the structure of 
extended sources and to search for variability of the individual 
objects (Ott, Eckart, Genzel 1999).
Due to the short time difference between the speckle imaging and the
spectroscopic measurements we know the brightness and exact positions 
of  all the prominent Sgr~A$^*$ cluster members that contribute to the 
observed flux in the slit we used during our spectroscopic measurements.
In Fig.4 we show our new proper motion determination of the 
fastest moving source 
S1 close to the position of the compact radio source Sgr~A$^*$.

\begin{figure}
\resizebox{\hsize}{!}{\includegraphics{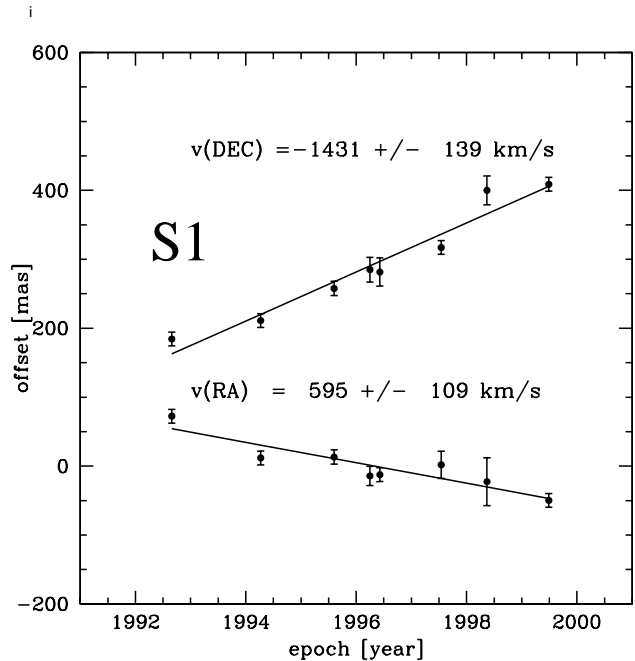}}
\hfill
\caption{
Proper motion of the fast moving object S1 close to the position
of the compact radio source Sgr~A$^*$. The plot includes all
data between 1992 and 1999. The declination graph 
has been moved upwards by 300~mas.
}
\label{fig4}
\end{figure}

\section{Conclusions}

We have presented new high angular and high spectral resolution data
obtained with the infrared spectrometer ISAAC at the VLT UT~1 and SHARP
at the NTT.
Our new spectra prove the lack of strong CO bandhead absorption
on the fast moving stars in the direct vicinity of the compact
radio source Sgr~A$^*$.
The new results are  fully consistent with the
most recent work on the central mass distribution (Genzel et al. 1999, 
Ghez et al. 1998, Eckart and Genzel 1996, 1997, Genzel et al. 1997, Eckart
and Genzel 1997). These new results further strengthen 
the very convincing case for a large compact mass at
the center of the Galaxy and 
indicate that its most likely current configuration 
is an inactive black hole (Maoz 1998).

\vskip 0.3cm 
\noindent
{\bf {\it Acknowledgements: \/}}
We thank A. Ghez, M. Morris, S.R. Stolovy, and M. Lehnert 
for interesting discussions and comments.
We are also grateful to the VLT UT1- and the NTT-team and especially 
to U.Weidenmann and H.Gemperlein for their interest and technical 
support of SHARP at the NTT.

\end{document}